\documentclass[useAMS,usenatbib]{mn2e}
\usepackage{epsfig}
\usepackage{times}
\usepackage{amssymb}

\title[Constraints on $\sigma_8$ from simulations]{Constraints on
  $\sigma_8$ from galaxy clustering in $N$-body simulations and
  semi-analytic models}

\author[G. Harker et al.]{Geraint Harker,$^{1,2}$\thanks{E-mail:
  harker@astro.rug.nl} Shaun Cole$^{1}$ and Adrian Jenkins$^{1}$\\
  $^{1}$Department of Physics, University of Durham, Science Laboratories,
  South Road, Durham DH1 3LE\\
  $^{2}$Kapteyn Astronomical Institute, University of Groningen, PO
  Box 800, 9700AV Groningen, the Netherlands}

\begin{document}

\date{\today}

\maketitle

\begin{abstract}
We generate mock galaxy catalogues for a grid of different
cosmologies, using rescaled $N$-body simulations in tandem with a
semi-analytic model run using consistent parameters. Because we
predict the galaxy bias, rather than fitting it as a nuisance
parameter, we obtain an almost pure constraint on $\sigma_8$ by
comparing the projected two-point correlation function we obtain to
that from the SDSS. A systematic error arises because different
semi-analytic modelling assumptions allow us to fit the $r$-band
luminosity function equally well. Combining our estimate of the error
from this source with the statistical error, we find $\sigma_8=0.97\pm
0.06$. We obtain consistent results if we use galaxy samples with a
different magnitude threshold, or if we select galaxies by $b_J$-band
rather than $r$-band luminosity and compare to data from the
2dFGRS. Our estimate for $\sigma_8$ is higher than that obtained for
other analyses of galaxy data alone, and we attempt to find the source
of this difference. We note that in any case, galaxy clustering data
provide a very stringent constraint on galaxy formation models.
\end{abstract}


\section{Introduction}

For a given set of cosmological parameters in $\Lambda$CDM, the
clustering of dark matter can be studied very accurately through
$N$-body simulations \citep[e.g.,][]{SPR05b}, or for that matter,
through analytic models calibrated by simulations
\citep[e.g.,][]{SMI03}. The clustering of dark matter is not usually
observed directly, however, though weak lensing shear-shear
correlations can provide (at present noisy) estimates.  Redshift
surveys may furnish us with galaxy clustering statistics \citep[see,
e.g.,][]{PEA03}, while weak lensing measurements, for example,
normally probe the cross-correlation between galaxies and dark matter
\citep[and references therein]{REF03}.

Galaxy clustering statistics derive a great deal of their power to
constrain cosmological parameters by constraining the scale at which
the power spectrum `turns over' on large scales, which complements the
high-redshift CMB constraint on this scale rather well. The baryonic
features in the correlation function or power spectrum add to the
effectiveness of the constraint \citep{COL05,EIS05}. The scales used
in these joint constraints tend to be large scales, where the
evolution of clustering is still in the linear regime or where
deviations from linearity can be more readily modelled. Moreover, in
this regime the galaxy correlation function is expected, in the
absence of non-local effects, to have the same shape as the mass
correlation function, though offset by a constant factor \citep[see,
e.g.,][]{COL93}. This offset -- the (square of the) bias -- depends on
the galaxy population under consideration; it depends, for example, on
the threshold luminosity of the sample. Because of this uncertainty,
when the galaxy correlation function is used to constrain cosmology,
information on its overall normalization is not normally used, and the
constraints come entirely from its shape.

In this paper we generate synthetic galaxy clustering statistics by
painting galaxies from a semi-analytic model onto dark matter
distributions given by $N$-body simulations.  We then compare these
clustering statistics with those from the SDSS to attempt to constrain
cosmology.  We can see the possibility for various benefits from our
approach. Firstly, because we attempt to generate realistic catalogues
with full galaxy properties, we can make a \emph{prediction} for the
bias factor of a given galaxy sample and hence use the overall
normalization of the correlation function in our cosmological
constraints. In particular, we may be able to constrain $\sigma_8$,
which is not possible for normal techniques employing galaxy
clustering. Secondly, because we populate the simulations on a
halo-by-halo basis rather than just assuming that galaxies
approximately trace mass on large scales, we generate a theoretical
prediction for the small-scale, nonlinear clustering.  We can
therefore attempt to use this information in our cosmological
constraints too.

Our constraints are largely independent from those using the CMB, and
involve different assumptions (though we consider only flat models,
which one could regard as implicitly using CMB results). Because dark
energy has an effect on structure formation, and different forms of
dark energy might affect it in different ways at late times, it is
useful to have an independent, low-redshift constraint on $\sigma_8$
that does not rely on a joint analysis with high-redshift data
\citep*{DOR01,BAR06}. A joint analysis would tend to be more model-dependent
as one must be able to model what happens in the gap between observed
snapshots of the Universe.

\section{Constructing galaxy catalogues}

\subsection{Simulations}\label{subsec:sims}

In choosing parameters for our simulations, our aim was to generate
simulation outputs for a range of different values of
$\Omega_\mathrm{M}$ and $\sigma_8$ so that we could examine galaxy
clustering as a function of these quantities. Given our focus on
constraining $\sigma_8$, we opted to generate outputs with $\sigma_8$
taking values between 0.65 and 1.05, regularly spaced in steps of 0.05.

\begin{figure}
  \begin{center}
    \leavevmode
    \psfig{file=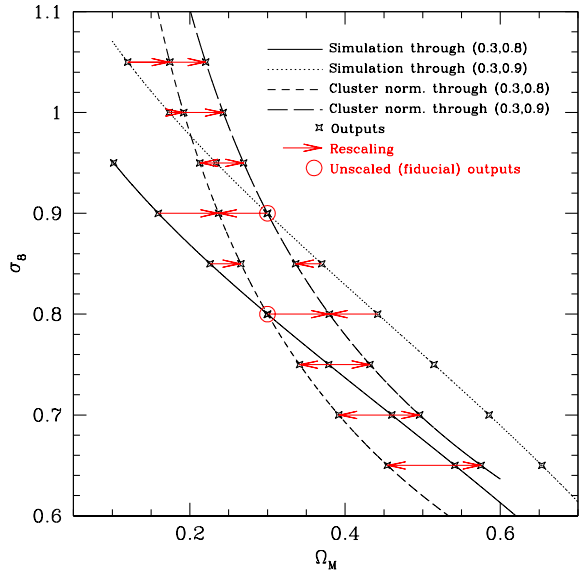,width=8cm}
    \caption{The position in the $(\Omega_\mathrm{M},\sigma_8)$ plane
      of the outputs of our two simulations. The line connecting the
      outputs shows how the values of $\Omega_\mathrm{M}$ and
      $\sigma_8$ change as the simulation evolves, where we track the
      instantaneous values of these parameters rather than the values
      they will have at the final time, which would be the more
      conventional labelling (and would, of course, not change during
      the course of the simulation). The solid line corresponds to
      Run~1 and the dotted line corresponds to Run~2. Low redshift
      outputs (lower density, more clustered) are in the top left,
      while high redshift outputs (higher density, less clustered) are
      in the bottom right. Also shown are the curves described by the
      cluster normalization condition that
      $\sigma_8\Omega_\mathrm{M}^{0.5}=\mathrm{const.}$ for two
      different values of this constant. We rescale simulation outputs
      so that they lie on these curves. The rescaling is shown
      schematically by the red arrows.}\label{fig:outplot}
  \end{center}
\end{figure}

Measurements of the abundance of clusters constrain the high-mass end
of the halo mass function, and hence constrain a combination of
$\Omega_\mathrm{M}$ and $\sigma_8$ \citep*[e.g.,][]{EKE96}. This
combination is, very approximately,
$\sigma_8\Omega_\mathrm{M}^{0.5}$. To test if we could break this
degeneracy, we have generated two grids of models. For Grid~1, the
parameters of each model satisfy
$\sigma_8\Omega_\mathrm{M}^{0.5}=0.8(0.3)^{0.5}$, while for Grid~2
they satisfy $\sigma_8\Omega_\mathrm{M}^{0.5}=0.9(0.3)^{0.5}$. Within
each grid, $\sigma_8$ takes on its full range of values between $0.65$
and $1.05$. It would be very difficult to distinguish between two
cosmologies lying on the same grid using cluster abundances. The two
`cluster normalization' curves, with
$\sigma_8\Omega_\mathrm{M}^{0.5}=\mathrm{const.}$, are shown as the
long-dashed and short-dashed lines in Fig.~\ref{fig:outplot}. The
pairs $(\Omega_\mathrm{M},\sigma_8)$ labelling the cosmologies we
analyse are plotted as crosses on these curves.

We extract the mass distribution for these cosmologies from two
simulations run using the \textsc{tree-pm} $N$-body code
\textsc{gadget2} \citep*{SPR01b,SPR05a}.  Each simulation follows the
evolution of $512^3$ particles in a $300\ h^{-1}\ \mathrm{Mpc}$
box. We have stored the simulation output at several redshifts. Each
of these snapshots of the mass distribution is then interpreted as a
$z=0.1$ snapshot of a simulation with a different cosmology, to avoid
having to run a great number of simulations.  We choose $z=0.1$ since
this is near the median redshift of the main SDSS and 2dFGRS galaxy
samples.  The output redshifts are chosen so that once the simulations
are relabelled as $z=0.1$ snapshots, the value of $\sigma_8$ at $z=0$
for each simulation falls onto a regular grid.  Each simulation then
gives us snapshots with $\sigma_8$ taking values between 0.65 and
1.05, regularly spaced by 0.05.  Table~\ref{tab:outputs} gives the
value of $\Omega_\mathrm{M}$ and $\sigma_8$ in these relabelled
snapshots. We have chosen the simulation parameters such that the
first simulation, `Run 1', has $\Omega_\mathrm{M}=0.3$ at its
$\sigma_8=0.8$ output, while the second simulation, `Run 2', has
$\Omega_\mathrm{M}=0.3$ at its $\sigma_8=0.9$ output.  When we perform
a further rescaling of $\Omega_\mathrm{M}$ (see below) it is these
central snapshots which remain unchanged.  The initial conditions are
calculated using a \citet{BBKS} power spectrum with shape parameter
$\Gamma=0.14$ and with primordial spectral index
$n_\mathrm{s}=1$.  A smooth power spectrum was most convenient in the
light of the rescalings we carry out on the final output, but in fact
the \citet{BBKS} power spectrum with $\Gamma=0.14$ was found to be a
good fit to the \textsc{cmbfast} \citep{SEL96} spectrum with
$\Omega_\mathrm{b}=0.045$ used for the Millennium Simulation
\citep{SPR05b}, the parameters of which were chosen to be in agreement
with the one-year WMAP results \citep{SPE03}.

Once we have relabelled the simulation snapshots as $z=0.1$ snapshots,
they lie on a curve in $(\Omega_\mathrm{M},\sigma_8)$ space which
reflects the way the dark matter density is reduced and the amplitude
of clustering is increased as the simulation evolves. These curves are
shown as the solid and dotted lines in Fig.~\ref{fig:outplot}. We
rescale $\Omega_\mathrm{M}$ in each snapshot, so that instead the
snapshots lie on one of the cluster-normalized curves described
above. The rescaling is achieved in practice by applying the results
of \citet{ZHE02}.  If the particle mass is scaled in the obvious way
to obtain the desired $\Omega_\mathrm{M}$, the particle velocities
must also be scaled to compensate, else the haloes no longer satisfy
the virial relation between their kinetic and potential energy, and
the galaxy populations of haloes are easily distinguished in the
different cosmologies via dynamical observables. The rescalings
in $\Omega_\mathrm{M}$ which move a snapshot onto the
cluster-normalization curve are shown schematically as red arrows in
Fig.~\ref{fig:outplot}. Each cluster-normalized grid contains rescaled
snapshots from both simulations, and the simulation parameters were
chosen so that the rescalings would never have to be too large. For
some of the cosmologies on our grid, we could choose to rescale from
either of our simulation runs without having to change
$\Omega_\mathrm{M}$ by a large factor. We have used these cosmologies
to test that the results using either simulation run are consistent,
and hence that our rescaling works as expected.

\begin{table}
  \caption{Cosmological parameters of simulation outputs after having
  been relabelled as \hbox{$z=0.1$} outputs. We follow the usual convention that these are the parameter values the simulation would have if evolved further to $z=0$.}
  \label{tab:outputs}
  \begin{center}
  \begin{tabular}{ccccc}
    \cline{1-2}\cline{4-5}
    \multicolumn{2}{c}{Run 1}&&\multicolumn{2}{c}{Run 2}\\
    $\Omega_\mathrm{M}$&$\sigma_8$&&$\Omega_\mathrm{M}$&$\sigma_8$\\
    \cline{1-2}\cline{4-5}
    &&&0.120&1.050\\
    &&&0.173&1.000\\
    0.102&0.950&&0.234&0.950\\
    0.159&0.900&&0.300&0.900\\
    0.226&0.850&&0.370&0.850\\
    0.300&0.800&&0.442&0.800\\
    0.379&0.750&&0.514&0.750\\
    0.460&0.700&&0.585&0.700\\
    0.541&0.650&&0.654&0.650\\
    \cline{1-2}\cline{4-5}
  \end{tabular}
  \end{center}
\end{table}

The simulation code runs \textsc{subfind} \citep{SPR01a} on the fly,
providing us with a list of friends-of-friends haloes \citep{DAV85} of
more than 20 particles, and their substructures. We use a linking
length of $0.2$ times the mean inter-particle separation in the
friends-of-friends algorithm to identify the haloes. \textsc{subfind}
also allows us to identify the particle in the halo with the least
gravitational potential energy, which we use in our galaxy placement
scheme.

\subsection{Semi-analytic model}\label{subsec:sams}

The properties of galaxies in our catalogues are generated using the
semi-analytic galaxy formation code, \textsc{galform}
\citep{COL00,BEN02,BEN03,BAU05}. For the purposes of this paper, we
may consider a semi-analytic model as being a means of predicting,
given some dark matter halo at a redshift of interest, the galaxy
population of that halo.  Having that information, we can construct
galaxy luminosity functions, correlation functions, etc.\ that might
be considered the results or predictions of the model.

The first step in predicting the galaxy population of a halo is
calculating the merger history of the halo.  In simulations of
sufficiently high resolution and with a sufficiently large number of
outputs, this can be extracted from the $N$-body data. In the case of
\textsc{galform}, this has been done recently by \citet{BOW06} with
the Millennium Simulation \citep{SPR05b}; the same simulation has also
been used by \citet{CRO06} and \citet{LUC06} to generate catalogues
using a different semi-analytic code. The simulations we describe
above, by contrast, do not have sufficient resolution for us to
extract reliable merger trees for the haloes of interest.  A Monte
Carlo scheme based on the work of \citet{LAC93} and using the
algorithm described by \citet{COL00} is employed instead,
therefore. This generates a merger tree for a halo based only on the
halo mass, the cosmology and the initial power spectrum, and does not
use other data from the simulation. This scheme does not,
therefore, provide galaxy positions; our method of placing galaxies is
given, instead, in Section~\ref{subsec:galplace}.

Unfortunately, the statistical properties of merger histories generated by
this algorithm are not identical to histories extracted directly from
an $N$-body simulation \citep{COL07}.  \citet*{PAR07} and
\citet{NEI07} have devised empirically motivated modifications to the
algorithm to allow Monte Carlo trees to fit the simulation data
better. A detailed analysis of the effect of such a modification on
semi-analytic galaxy properties is beyond the scope of this paper. We
have, though, tested some of our results using the new algorithm of
\citet{PAR07}, and find that for our purposes the new trees make
little difference.

Given the merger history of a halo, the model computes the evolution
of the baryonic content of the halo using a variety of analytic
prescriptions.  Many of the equations governing the physical processes
modelled by \textsc{galform} contain parameters which may be
adjusted. Some of these (for example, the form factor
$f_\mathrm{orbit}/c$ which governs the size of merger remnants) have
a `natural' value determined by the physics; others (those
governing the angular momentum distribution of infalling haloes, say)
are derived by comparison to more detailed simulations. The function
of allowing these parameters to change, then, is to allow
investigation into the magnitude of the effect of different physical
processes on the resulting galaxy properties in the model. Other
parameters have no natural value, and can only be fixed by requiring
that they take values which allow the model to fit observations. Much
of the time, if we are able to fit some set of observations
satisfactorily by choosing the parameters of the model judiciously,
the same set of observations could also be fit reasonably well by some
very different choice of parameters. Therefore, within the
\textsc{galform} framework, we have different models using different
physics which are equally good at matching the observations (though
this may not, of course, be the case if we were to choose a different
set of observations to constrain the model).

Our aim here is to try to constrain cosmological parameters by
comparing clustering statistics from a simulation populated with
semi-analytic galaxies to the corresponding measurements in an
observational survey. We would hope that our constraints are
insensitive to the precise semi-analytic model used, and we would like
to test whether this is the case. Therefore, although we use only one
code, \textsc{galform}, we use three different `models', in the sense
of different combinations of the physics we attempt to model and the
parameters governing that physics. In the remainder of this section of
the paper, we discuss the technical differences between the three
models before briefly describing how galaxies are placed in the
simulations in Section~\ref{subsec:galplace}. A reader uninterested
in the details of the models may therefore wish to skip to
\ref{subsec:galplace}, or to our results in Section~\ref{sec:res}. The
three models are as follows:
\begin{itemize}
\item The fiducial model of \citet{COL00}. This is successful in
  matching several sets of observations, including the $B$- and
  $K$-band luminosity functions, galaxy colours and mass-to-light
  ratios for galaxies of different morphologies, the cold gas mass in
  galaxies, galaxy disc sizes and the slope and scatter of the
  $I$-band Tully-Fisher relation. Unfortunately, though, it assumes a
  cosmic baryon fraction, $\Omega_\mathrm{b}$, of only 0.02. This is
  inconsistent with recent estimates from Big Bang nucleosynthesis
  \citep[e.g.,][]{OME01} and the cosmic microwave background
  \citep{SPE07}. Nevertheless, we feel it is worthwhile to include
  this model in our analysis as a well recognized and well understood
  model that has been thoroughly described and studied. In our
  figures, lines corresponding to output from this model are given the
  label `Cole2000'.
\item A model similar to the first, but with $\Omega_\mathrm{b}=0.04$,
  which is closer to current estimates. Since there are twice as many
  baryons as in the first model, if we leave the rest of the
  parameters unchanged then, as expected, the model is unable to match
  observations such as the luminosity function. Therefore we introduce
  a new physical process: thermal conduction in massive haloes (this
  is analysed in greater detail by \citealt{BEN03}). We simply assume
  that gas is unable to cool if the halo circular velocity,
  $V_\mathrm{circ}$, satisfies
  \begin{equation}
    V_\mathrm{circ}>V_\mathrm{cond}\sqrt{1+z}\quad ,
  \end{equation}
  where $V_\mathrm{cond}$ is a parameter we may adjust. This suppresses
  the problematic bright end of the luminosity function; the effect is
  similar, in fact, to more recent and more physically motivated
  implementations of feedback from active galactic nuclei in
  \textsc{galform} \citep{BOW06}. Though it is clearly rather crude,
  note that our objective here is only to produce a realistic enough
  galaxy catalogue to compare to observations. We are trying to mimic
  the effect of whatever physical process suppresses the bright end of
  the luminosity function, without having to adopt a complicated
  parametrization that is no better physically motivated than a more
  simple and understandable one. The label we give to this model in our
  figures is `C2000hib' (where `hib' stands for `high baryon fraction').
\item Another model with $\Omega_\mathrm{b}=0.04$, but now
  incorporating `superwinds'. In this model it is postulated that a
  galaxy's cold gas is heated strongly enough for it to be expelled
  completely from the halo, rather than returning to the reservoir of
  hot gas associated with the halo. In fact, the model is derived from
  that used by \citet{BAU05} to reproduce the abundance of faint
  galaxies detected at submillimetre wavelengths. This also
  incorporates the additions and refinements to \textsc{galform}
  described by \citet{BEN03}. These include a modification to the
  assumed profile of the halo gas, a more sophisticated treatment of
  conduction, and a more detailed treatment of galaxy mergers, in
  particular the effects of tidal stripping and dynamical friction
  \citep{BEN02}. They also include a simple model of the effect of
  reionization on small haloes, where cooling is prevented if
  $V_\mathrm{circ}<V_\mathrm{cut}$ and $z<z_\mathrm{cut}$ for two
  parameters $V_\mathrm{cut}$ and $z_\mathrm{cut}$. The strength of
  superwind feedback is parametrized by $V_\mathrm{SW}$, the
  characteristic velocity of the wind. The model is denoted `Model M'
  in our figures.
\end{itemize}

We wish to run each of these models in many different cosmologies, in
order to generate galaxy catalogues in which the $N$-body component
and the semi-analytic component are consistent.  Changing cosmology
naturally changes the galaxy population predicted by each model,
however, so that even if we fix the parameters such that the galaxies
match observational constraints in one fiducial cosmology, they are
unlikely to match in other cosmologies. We therefore tweak the
parameters between different cosmologies to try to match the data. It
is not possible to do this for the full range of even the primary
constraints described by \citet{COL00}. We restrict ourselves to a
comparison with the $^{0.1}r$-band SDSS luminosity function of
\citet{BLA03b} at $z=0.1$. Even then, to make the problem tractable
and to ensure our three models remain distinct, we restrict the
parameters we allow ourselves to vary to:
\begin{itemize}
\item $V_\mathrm{SW}$, for Model M only.
\item $V_\mathrm{cond}$, for the C2000hib model only.
\item $V_\mathrm{cut}$, one of the parameters controlling
  reionization. Though we experimented with changing this for all the
  models, all the ones below have either $V_\mathrm{cut}=0$ (Cole2000
  and C2000hib) or $V_\mathrm{cut}=60$ (Model M). As well as being
  simpler, this also helps make Model M more distinct from the other
  two.
\item $V_\mathrm{hot}$ and $\alpha_\mathrm{hot}$. These are closely
  linked but we vary them independently. They control the strength of
  standard (i.e.\ not superwind) supernova feedback in the following
  way. The rate of change of the mass of hot gas and of cool gas in a
  halo are linked with the instantaneous star formation rate, $\psi$,
  by:
  \begin{equation}
    \dot{M}_\mathrm{hot}=-\dot{M}_\mathrm{cool}+\beta\psi
  \end{equation}
  \citep[equation 4.7]{COL00}. $\beta$ is related to the circular
  velocity of the galaxy disc, $V_\mathrm{disc}$, by
  \begin{equation}
    \beta=(V_\mathrm{disc}/V_\mathrm{hot})^{-\alpha_\mathrm{hot}}
  \end{equation}
  \citep[equation 4.15]{COL00}.
\end{itemize}

Some of the cosmologies below require quite extreme, perhaps
unphysical, parameter values. In some cases, \textsc{galform} is
reluctant to run, while in others the fit for some observations is
compromised in an attempt to fit the $^{0.1}r$-band luminosity
function well. In addition to running each model with tweaked
parameters in each cosmology, therefore, we also run each model in
each cosmology using the same parameters as the central cosmology of
Run~1, in which $(\Omega_\mathrm{M},\sigma_8)=(0.3,0.8)$.

We are not able to produce a good fit to the luminosity function in a
$\chi^2$ sense, even allowing these parameters to vary. This may be a
concern when comparing clustering statistics to observational
data. Volume-limited galaxy samples, to which we wish to compare our
results later, are chosen such that all the galaxies are brighter than
some given absolute magnitude limit. If we choose a sample of
semi-analytic galaxies with the same magnitude limit, then because our
luminosity function is wrong we may not be choosing a sample that
necessarily corresponds to the observational one, even within our
model. Therefore we instead select semi-analytic galaxy samples with a
magnitude such that the sample has the same space density as the
corresponding observational sample. This means that when we adjust the
parameters of the model to match the luminosity function, it is more
important for our purposes to match its overall \emph{shape} rather
than its magnitude normalization.

The $\Upsilon$ parameter of \citet{COL00} is related to this sort of
scaling of the luminosity function. It was introduced to account for
brown dwarfs, which absorb some of the mass of gas assumed to be tied
up in stars, but without producing light. It is defined by
\begin{equation}
\Upsilon=\frac{\textrm{mass in visible stars + mass in brown
    dwarfs}}{\textrm{mass in visible stars}}
\end{equation}
(\citealt{COL00}, equation~5.2). Clearly, then, we must have $\Upsilon
\geq 1$ given this physical explanation. The result of including this
effect is to scale luminosities by a factor $1/\Upsilon$. For each
\textsc{galform} model we run, we compare the resulting $^{0.1}r$-band
luminosity function with the observational value from the SDSS (in
fact, we compare only one point near the characteristic luminosity,
$L_\ast$). We express the difference between the two in terms of the
$\Upsilon$ parameter: the (reciprocal of the) amount by which we would
have to scale the luminosity of the semi-analytic galaxies to match
the data. Sometimes this requires $\Upsilon<1$. Therefore, when we
give a value of $\Upsilon$ below, it should be treated only as an
indication of the amount by which we would have to scale luminosities
so that when we select a galaxy sample by a number density threshold
then it would have the same luminosity threshold as the observational
sample. Note that we calculate $\Upsilon$ by reference to a specific
point on the SDSS $^{0.1}r$-band luminosity function. Its exact value
would change if we normalized at a different point (since the model
luminosity function is not the same shape as the observational one),
or in a different band (since the colour of model galaxies may be
incorrect).

Once we have given ourselves the freedom to scale the luminosity
function in this way, then, the effect of varying the parameters we
allow ourselves to change to try to match the shape of the luminosity
function is as follows:
\begin{itemize}
\item Increasing $V_\mathrm{SW}$, or decreasing $V_\mathrm{cond}$,
  tends to steepen the bright-end slope, i.e.\ give fewer very bright
  galaxies.  $V_\mathrm{SW}$ is only non-zero for Model M;
  $V_\mathrm{cond}$ is only finite for the C2000hib model.
\item Increasing $V_\mathrm{cut}$ reduces the slope at the faint end,
  reducing the number of the faintest of the galaxies we study.
\item Increasing $V_\mathrm{hot}$ tends to suppress the overall space
  density of galaxies. Because of the effect of the other parameters,
  it is most useful for adjusting the abundance of galaxies of around
  $L_\ast$, or a little fainter.
\item Changes in $\alpha_\mathrm{hot}$ can be viewed as modulating the
  effect of changing $V_\mathrm{hot}$. Visually, for typical values of
  $V_\mathrm{hot}$, increasing $\alpha_\mathrm{hot}$ flattens the
  faint-end slope, typically over a wider range of luminosities than
  $V_\mathrm{cut}$.
\end{itemize}
We usually find that to make the bright-end slope steeper and to make
the faint-end slope shallower, as required by the data, needs all
parameters tweaked to give larger amounts of feedback. This tends to
have the overall effect of reducing the predicted luminosities,
leading to $\Upsilon<1$ as mentioned above. Requiring $\Upsilon\geq 1$
would therefore require us to compromise one component or other of
the shape in these models. Since we later rescale to match space
densities anyway, we opt not to make this compromise. Given two
parameter combinations which both match the luminosity function
reasonably well and which both give $\Upsilon<1$, we use the
$\Upsilon$ parameter as a tie-breaker, selecting the combination which
gives $\Upsilon$ closer to unity. We have also checked that each model
is at least qualitatively consistent with the other primary
\textsc{galform} constraints (the Tully-Fisher relation, disc sizes,
morphological mix, metallicities and gas fractions -- see
\citealt{COL00}).

\begin{figure}
  \begin{center}
    \leavevmode
    \psfig{file=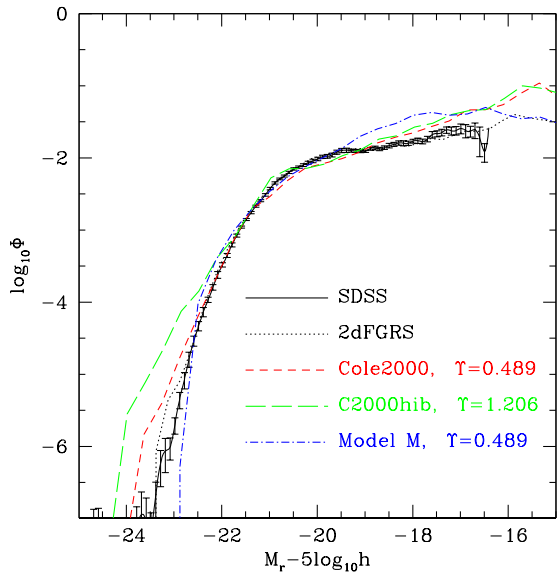,width=8cm}
    \caption{$r$-band luminosity functions for the three fiducial
    \textsc{galform} models, compared to the SDSS luminosity function
    of \citet{BLA03b} (solid line) and a 2dFGRS $r$-band
    luminosity function (dotted line, mostly obscured by the others) generated using the SuperCosmos $r_F$-band
    data \citep{HAM01}. The value of the scaling parameter $\Upsilon$, required to normalize the model luminosities for galaxies of a
    particular space density in this band, is also given in the
    legend. Errors on the SDSS luminosity function are only given for
    every tenth point, for clarity.}\label{fig:lfr}
  \end{center}
\end{figure}

The $r$-band luminosity function for each of our models in our
fiducial cosmology ($\Omega_\mathrm{M}=0.3$ and $\sigma_8=0.8$) is
shown in Fig.~\ref{fig:lfr}. Qualitatively, the agreement between the
models and the data is reasonable. The very sharp cutoff at the bright
end of the luminosity function in Model M is a generic feature of the
model. The lower space density of very faint galaxies in this model is
also generic, and comes from the introduction of reionization
(non-zero $V_\mathrm{cut}$). At first glance, it appears that the
Cole2000 model gives better agreement with the data at the bright end
than the C2000hib model, despite the inclusion of a feedback mechanism
specifically to solve this problem in the latter. Recall, though, that
the Cole2000 model has a lower baryon fraction, and despite this we
have to introduce relatively high levels of supernova feedback to
match the shape of the luminosity function. We therefore need
$\Upsilon\sim 0.5$ to recover the correct luminosities, while the
C2000hib model needs a much more physically palatable $\Upsilon\sim
1.2$. It may appear that our requirement for $\Upsilon \sim 0.5$ is
inconsistent with the original \citet{COL00} paper, the reference
model of which requires $\Upsilon = 1.38$. It is not inconsistent, for
a few reasons. Firstly, we use the label `Cole2000' for our model
because it uses equivalent code with the same physics governed by the
same parameters as the models of \citet{COL00}. As we have just noted,
however, some of the parameters take different values in our fiducial
model in order to try to match the shape of the $r$-band luminosity
function. Secondly, while the reference model of \citet{COL00} had
$\sigma_8=0.93$, ours has $\sigma_8=0.8$. Thirdly, their $\Upsilon$
was calculated by reference to the value of the observed $b_J$ band
luminosity function at $L_\ast$ (though in fact the same correction
also provided a good match to the $K$-band luminosity function). Ours
is calculated by reference to the $r$-band luminosity function. We
match to a point slightly brighter than $L_\ast$ (where the
exponential cutoff has started to bite more deeply and the galaxies
are less abundant; the point at $M_r\sim-21.5$ can be seen quite
easily in Fig.~\ref{fig:lfr} as being where all the lines cross) since
we otherwise had problems calculating $\Upsilon$ for some of our
models with a very shallow faint-end slope and low galaxy number density.

\subsection{Galaxy placement}\label{subsec:galplace}

With the $N$-body simulations and the semi-analytic catalogues in
place, it remains to merge the two to create a synthetic galaxy
catalogue, or in other words to populate the simulations with our
galaxies. To each halo in the simulation we assign a semi-analytic
galaxy population for a random merger tree of the same mass. We then
place the central galaxy at the position of the particle with least
gravitational potential energy, and place the satellite galaxies on
random particles within the halo.

One might worry that given the resolution of our simulations, it is
possible for the semi-analytic model to predict that a halo in a
simulation contains a bright enough galaxy to enter our sample even
though the halo is not resolved with at least 20 particles, which is
our normal criterion for considering the halo to be resolved. To take
account of these galaxies, we calculate the number of such haloes
expected for the simulation volume for the \citet{JEN01} mass
function. We then take the galaxy populations predicted by
\textsc{galform} for these haloes and place the galaxies on random
particles in the simulation which are not in haloes. We do not expect
this to have a significant effect on clustering statistics, since
almost all galaxies which would be placed in unresolved haloes are
very faint, and in any case the halo bias as a function of mass is not
strongly varying in this regime \citep{COL89,MO99} so that we do not
lose too much accuracy by placing galaxies in haloes of the wrong
mass. None the less, we have checked that employing this scheme has
only a small effect on our measured correlation functions. Changing
the minimum resolved mass from 20 particles to 50 particles has only a
very small effect on the correlation function, as does ignoring the
`unresolved' galaxies entirely, even for a conservative mass limit of
50 particles. Moreover, this remains true even for galaxy samples
which are rather faint when compared to the magnitude limit of the
SDSS samples we will be considering, and which therefore provide a
more stringent test.

\section{Results}\label{sec:res}

\subsection{Observational samples}

\begin{figure*}
  \begin{center}
    \leavevmode
    \psfig{file=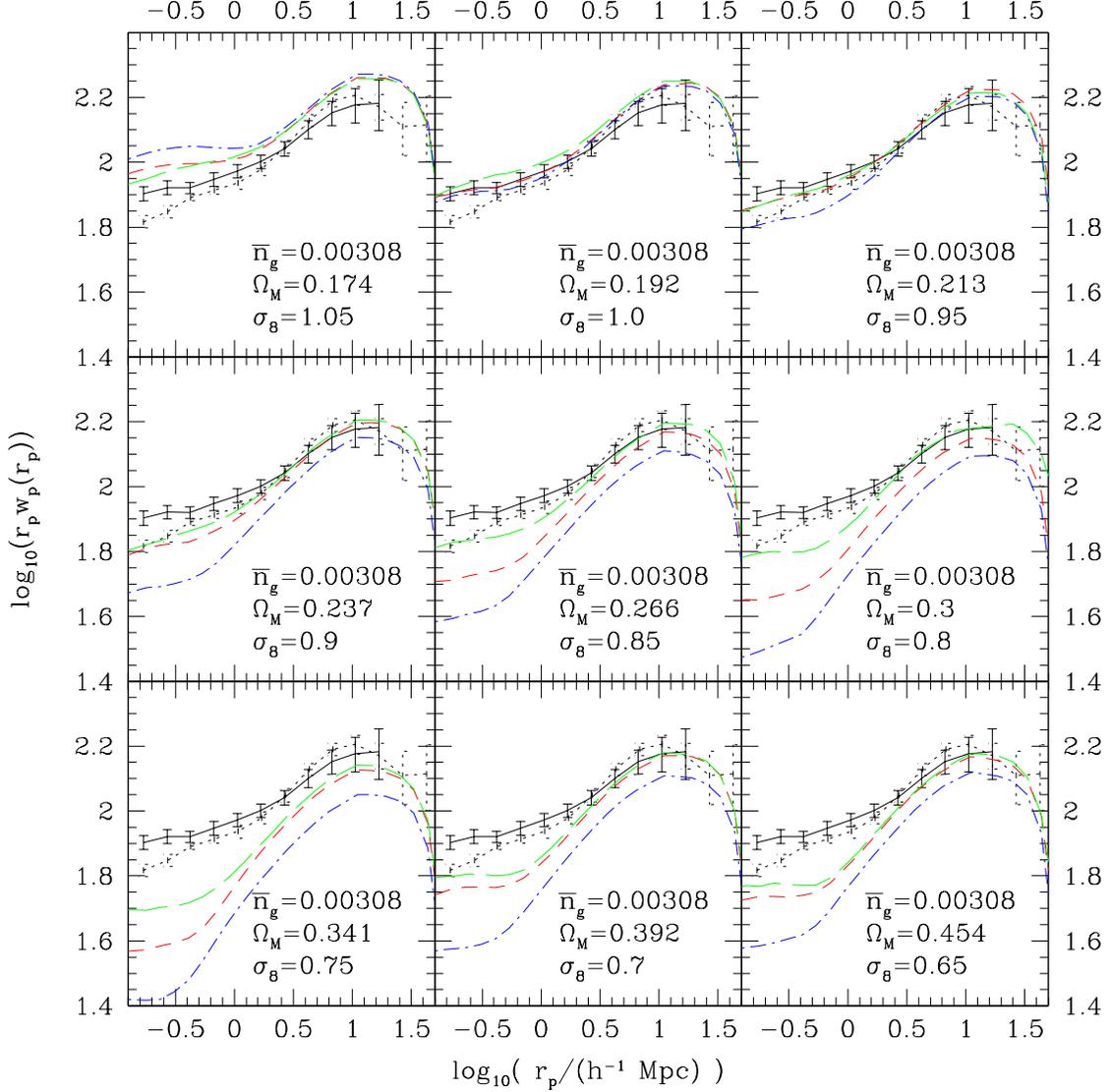,width=16cm}
    \caption{Clustering in our models and in the
    SDSS. $r_\mathrm{p}w_\mathrm{p}(r_\mathrm{p})$ is plotted for
    clarity. The solid, black line with error bars is the SDSS data;
    the dotted line shows the SDSS flux-limited sample for
    comparison. The coloured lines are from our three models:
    short-dashed red for Cole2000, long-dashed green for C2000hib and
    dot-dashed blue for Model M, as in Fig.~\ref{fig:lfr}. The nine
    different cosmologies form a grid in $\sigma_8$ with the
    cosmological parameters lying on the curve
    $\sigma_8\Omega_\mathrm{M}^{0.5}=0.8(0.3)^{0.5}$. This plot shows
    models for which we allow ourselves to tweak the \textsc{galform}
    parameters to match the luminosity function.}\label{fig:rwgrid1}
  \end{center}
\end{figure*}

In their study of the luminosity and colour dependence of the galaxy
correlation function using the main galaxy sample of the SDSS,
\citet{ZEH05} calculated the projected two-point correlation function
$w_\mathrm{p}(r_\mathrm{p})$ for ten different galaxy samples defined
by thresholds in absolute magnitude. We have been provided with these
correlation functions, and their covariance matrices calculated by
jackknife resampling. \citet{ZEH05} also tabulate the space density of
each sample, so it is straightforward for us to select corresponding
samples of semi-analytic galaxies.

Our cosmological constraints will use samples with a galaxy space
density $\bar{n}_g=0.00308\ h^3\ \mathrm{Mpc}^{-3}$, corresponding to
galaxies with $M_r-5\log_{10}h<-20.5$ in the SDSS. Our semi-analytic
catalogues have approximately twice the effective volume of the
observational sample, so when calculating how well our models fit the
data we use only the covariance matrix of the observational
correlation function to compute our errors, neglecting the statistical
errors on the simulated function. We use the sample of this space
density since it provides a good compromise between volume and space
density, giving relatively small errors, and since most of the
constraining power then comes from the galaxies of intermediate
luminosity which are modelled best by the semi-analytic code. We will,
though, briefly discuss the effect of using samples of a different
space density or selected in a different waveband.

\subsection{Constraints}\label{subsec:const}

\begin{figure}
  \begin{center}
    \leavevmode
    \psfig{file=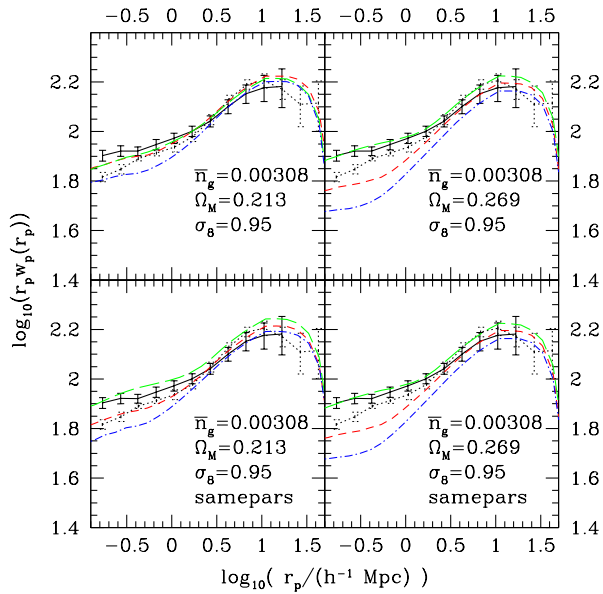,width=8cm}
    \caption{A comparison between our four sets of models for one
    particular value of $\sigma_8$. The top two panels are for models
    in which we tweak the semi-analytic parameters to match the
    $r$-band luminosity function, while for the bottom two the
    parameters stay the same as for our fiducial model.  The two
    left-hand panels are for the low-$\Omega_\mathrm{M}$ cluster
    normalization curve while the others are for the
    high-$\Omega_\mathrm{M}$ case. The colour coding and line styles
    of the models are the same as for Figs.~\ref{fig:lfr}
    and~\ref{fig:rwgrid1}.}\label{fig:rw_four}
  \end{center}
\end{figure}

We compare the clustering in our synthetic catalogues and in the SDSS
in Fig.~\ref{fig:rwgrid1}. We plot the quantity
$r_\mathrm{p}w_\mathrm{p}(r_\mathrm{p})$ since this scales out much of
the $r_\mathrm{p}$ dependence and makes differences in shape easier to
see. Fig.~\ref{fig:rwgrid1} shows our results for a grid of nine
cosmologies spaced regularly in $\sigma_8$ such that they lie on the
curve $\sigma_8\Omega_\mathrm{M}^{0.5}=0.8(0.3)^{0.5}$ (`Grid 1'). For
this plot we show the models for which we allow the semi-analytic
parameters to vary so as to provide a good match for the $r$-band
luminosity function.  Note that we have three other similar sets of
models: one which includes the same cosmologies but in which the
semi-analytic parameters are identical in each cosmology, and two more
in which the cosmologies lie on the same grid in $\sigma_8$ but which
have lower $\Omega_\mathrm{M}$, such that
$\sigma_8\Omega_\mathrm{M}^{0.5}=0.9(0.3)^{0.5}$ (`Grid 2'). In one
low-$\Omega_\mathrm{M}$ sequence the semi-analytic parameters are
allowed to vary, and in the other they are not. A figure similar to
Fig.~\ref{fig:rwgrid1} could be made for each of the latter three sets
of catalogues, but since the features turn out to be qualitatively
similar we do not show such plots here. We do, though, compare the
four sets for one particular value of $\sigma_8$ in
Fig.~\ref{fig:rw_four}. For each grid of cosmologies and for each
choice as to whether to allow the parameters to vary we have
catalogues for each of our three models, and therefore in total we
have twelve sets of catalogues each of which has nine members lying on
a regular grid in $\sigma_8$. The key to our numbering of these sets
is given in Table~\ref{tab:modelkey}.

Examining Fig.~\ref{fig:rwgrid1}, it is clear that some of our
catalogues fit the data better than others. For the higher $\sigma_8$
cosmologies, the shape of the models fits that of the data rather
well. The trend between cosmologies is consistent between the three
\textsc{galform} models: a higher amplitude of clustering for higher
$\sigma_8$, as expected. There are differences between the models,
however, especially on small scales, which must arise from differences
in the details of the halo occupation distribution predicted by the
models.

\begin{table}
\caption{The key to the model numbering
  used in Fig.~\ref{fig:bottomline}. The first column gives the
  label we assign to each of our 12 sets of populated simulations
  (each of which has nine cosmologies, regularly spaced in
  $\sigma_8$). The `Grid' in the second column refers to whether the
  cosmologies lie on the cluster normalized curve with high
  $\Omega_\mathrm{M}$ (Grid 1) or low $\Omega_\mathrm{M}$ (Grid
  2). The third column shows whether we adopt different
  parameters in different cosmologies, or whether they stay the same,
  while the fourth gives the \textsc{galform} model in use.}
\label{tab:modelkey}
\begin{center}
\leavevmode
\begin{tabular}{cccc}
\hline
Model no. & Grid & Same/diff. pars. & \textsc{galform} model\\
\hline
    1 & 1 & diff. & C2000hib \\
    2 & 1 & diff. & Cole2000 \\
    3 & 1 & diff. & M        \\
    4 & 1 & same  & C2000hib \\
    5 & 1 & same  & Cole2000 \\
    6 & 1 & same  & M        \\
    7 & 2 & diff. & C2000hib \\
    8 & 2 & diff. & Cole2000 \\
    9 & 2 & diff. & M        \\
    10 & 2 & same & C2000hib \\
    11 & 2 & same & Cole2000 \\
    12 & 2 & same & M        \\
\hline
\end{tabular}
\end{center}
\end{table}

\begin{figure}
  \begin{center}
    \leavevmode
    \psfig{file=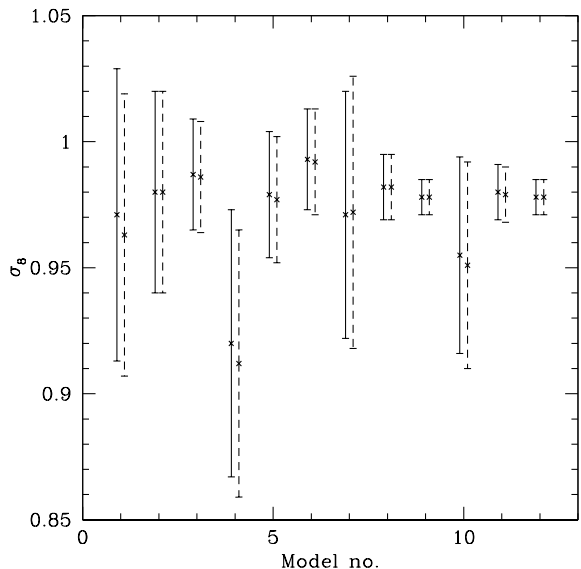,width=8cm}
    \caption{Constraints on $\sigma_8$. The $x$-axis shows the model
    number, the key to which is given in Table~\ref{tab:modelkey}. The
    $y$-axis shows the 1$\sigma$ constraint on $\sigma_8$ achieved in
    that particular model. The points with solid error bars are for
    the unmodified catalogues. The points with dashed error bars show
    how the constraints change when we make an empirical correction for the effect of halo assembly bias \citep*[and references therein]{CRO07}. We do this by modifying the correlation function
    according to the scale-dependent bias between shuffled and
    unshuffled \textsc{galform} catalogues in the Millennium
    Simulation, as described in Section~\ref{subsubsec:hab}.}\label{fig:bottomline}
  \end{center}
\end{figure}

The variation in the predicted correlation function between
cosmologies, and the consistent trend between models, supports our
hope that comparison to the SDSS correlation function can constrain
$\sigma_8$. For each set of nine cosmologies, we calculate $\chi^2$
with respect to the observed correlation function and its covariance
matrix, then fit a quadratic through the three points around the
minimum to interpolate and estimate the best-fitting $\sigma_8$ and
its 1-$\sigma$ error. The result of applying this procedure is given
in Fig.~\ref{fig:bottomline}. There, we give an estimate of $\sigma_8$
and its errors for each of our twelve different sets of populated
simulations, as the black crosses and error bars. The model number
referred to on the $x$-axis is explained in Table~\ref{tab:modelkey}.

A few comments may be made about Fig.~\ref{fig:bottomline}. Firstly,
some of the sets of models yielded no value of $\sigma_8$ for which
the simulated correlation function was an acceptable fit to the
observed one. The large $\chi^2$ and $\Delta\chi^2$ values then result
in a spuriously small error bar. This is the case for models 9, 11 and
12, so the constraints coming from those models should be
ignored. Secondly, recall that we ran only two $N$-body simulations,
using only the outputs given in Table~\ref{tab:outputs}. In fact,
while one was run with a larger value for $\sigma_8$ (and therefore a
smaller $\Omega_\mathrm{M}$ for an output with given $\sigma_8$), it
was started with initial conditions where the different Fourier modes
of the density field were given the same phase as in the
low-$\sigma_8$ simulation. This means that the constraints from the
different sets of synthetic galaxy catalogues are not independent, but
should be used to give an indication of the systematic error arising
from the choice of semi-analytic model parameters and the assumed
value of $\Omega_\mathrm{M}$ (which we do not constrain). Note that we
have also plotted two constraints for each model in
Fig.~\ref{fig:bottomline}. On the left, with solid error bars, are the
constraints derived just as we have described. We discuss the
estimates on the right, with dashed error bars, in
Section~\ref{subsubsec:hab}.

\subsection{Other catalogues}

\subsubsection{Halo assembly bias}\label{subsubsec:hab}

The points with dashed error bars in Fig.~\ref{fig:bottomline} show a
constraint after we attempt to correct our simulated correlation
functions for the effect of so-called `halo assembly bias'. This is an
effect that may arise if one of the assumptions we make when
populating our simulation with galaxies is incorrect. We assume that
the distribution from which the properties of the galaxy content of a
halo are drawn depends only on the mass of the halo. In other words,
since the halo merger tree is the basic input to our semi-analytic
model, we assume that the distribution from which the properties of a
halo's merger tree are drawn depends only on halo mass. This is
explicitly the case for the Monte Carlo merger trees we use here,
since it is a result of the underlying extended Press-Schechter theory
\citep{PRE74,BOW91,BCEK,LAC93}. This result was also supported by work
on simulations \citep{LEM99}. More recently, however, the advent of
larger simulations with better resolution has allowed this result to
be challenged.  \citet*{GAO05} showed that old haloes of a given mass
in the Millennium Simulation were more strongly clustered than young
haloes of the same mass, while \citet{MCF_06} demonstrated that halo
formation time is a function of halo environment as well as halo mass
using an independent set of merger trees in the same
simulation. Because halo age is a property of the halo merger tree,
this shows that the assumption we have described is violated. In fact,
the variation of $N$-body merger trees with environment has been
studied in other large simulations \citep{MAU07}.  More generally,
this environmental dependence formally invalidates the straightforward
application of halo models of galaxy clustering
\citep{BEN00,SEL00,BER02,COO02}. It has stimulated theoretical
attempts to explain the departure from the extended Press-Schechter
prediction (e.g., \citealt{SAN07}; \citealt*{WAN07}), and to look for
possible effects on the galaxy population both observationally
\citep*{YAN06} and in models (\citealt{CRO07}; \citealt{ZHU06}).

\begin{figure*}
  \begin{center}
    \leavevmode
    \psfig{file=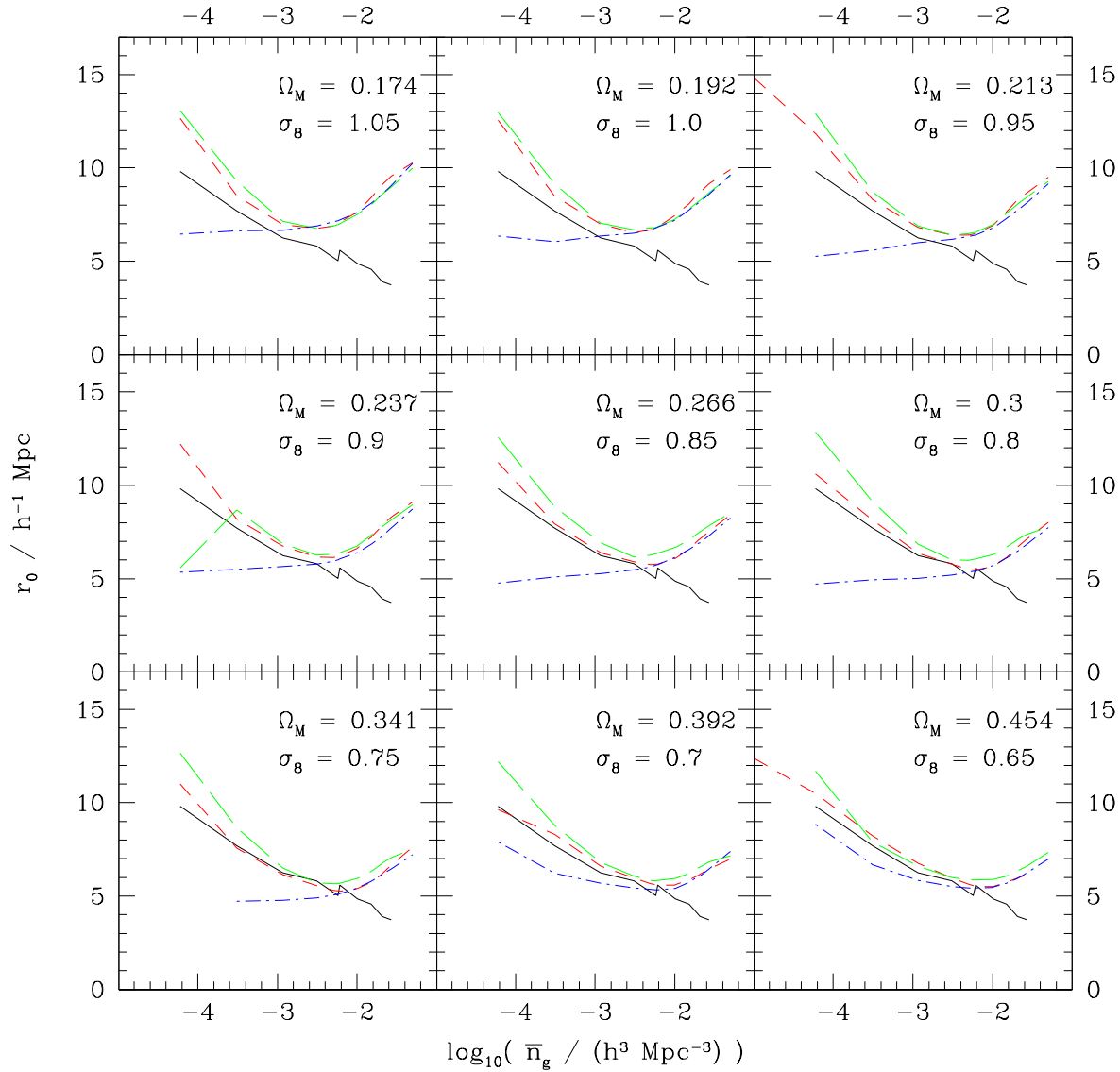,width=16cm}
    \caption{Correlation length as a function of sample space density
    for $r$-band selected galaxies, for the same set of models and
    with the same colour coding and
    line styles as Fig.~\ref{fig:rwgrid1}.}\label{fig:biasgrid1}
  \end{center}
\end{figure*}

Our correlation functions are corrected using a \textsc{galform}
catalogue generated in the Millennium Simulation. The version of
\textsc{galform} which generates the catalogue takes as its input the
actual $N$-body merger tree of each halo \citep{BOW06}. This catalogue
therefore incorporates environmentally dependent halo formation. In a
similar spirit to \citet{CRO07}, we shuffle this catalogue, assigning
to each halo the galaxy population of a random halo of the same
mass. This destroys any connection between the environment of a halo
of given mass and its galaxy population. We calculate the galaxy
correlation function for a range of different magnitude thresholds for
both the original and shuffled catalogues. This gives us an estimate
of the effect of halo assembly bias, for \textsc{galform} galaxies at
least: we note that while our results are qualitatively consistent
with those of \citet{CRO07}, the two semi-analytic models do not
respond identically. We have calculated the scale-dependent ratio
between the correlation functions, and then used this ratio (for a
sample of appropriate space density) to correct the correlation
functions used for our constraints. This is intended only to give an
estimate of the size of the systematic error on our constraints coming
from halo assembly bias. As Fig.~\ref{fig:bottomline} shows, the error
from this source is small in comparison to the statistical errors, for
our model at least.

\subsubsection{Luminosity dependence}\label{subsubsec:lumdep}

We calculate the correlation length of the samples by fitting a power
law to $\xi(r)$ for $2<r/(h^{-1}\ \mathrm{Mpc})<20$; that is, we
parametrize the correlation function as $\xi(r)=(r/r_0)^{-\gamma}$
where $r_0$ is the correlation length. We have done this for all our
samples of all luminosities, so we are able to plot the correlation
length as a function of sample space density (or, equivalently, as a
function of sample luminosity threshold) in
Fig.~\ref{fig:biasgrid1}. The black line in the plots shows the
corresponding result from the SDSS. The SDSS data show a steady
increase in clustering strength with luminosity (i.e.\ with decreasing
space density) apart from a feature at $\bar{n}_\mathrm{g}\approx
0.006\ h^3\ \mathrm{Mpc}^{-3}$ corresponding to the difference between
two $M_r^\mathrm{max}=-20.0$ samples: one has a large, overdense
region at $z\sim 0.08$ excised (see \citealt{ZEH05} for details) and
has lower space density but, as might be expected, weaker clustering
than the sample where this region is retained.

For many cosmologies, the Cole2000 model and the updated, higher
baryon fraction C2000hib model do a reasonable job of matching the
luminosity-dependent clustering in the SDSS, especially for samples of
moderate to low space density. Model M, which invokes superwinds (see
Section~\ref{subsec:sams}), does not do so well, predicting very
little luminosity dependence. This may be because the feedback effects
are so extreme in large haloes that their central galaxies are little
brighter (if at all) than those at the centre of less massive, less
biased haloes. The other two models tend to have the opposite problem
in the brightest samples: they tend to predict too high an amplitude
of clustering. This could be due to too tight a relationship between
halo mass and galaxy luminosity (perhaps because in reality feedback
is more efficient or more stochastic): none of the brightest galaxies
is scattered into lower mass, less biased haloes. A generic feature of
the \textsc{galform} models seems to be an upturn in the clustering
amplitude at high space density. This suggests that too many of the
faint galaxies generated by the model reside in high mass haloes. This
may be related to the fact that it is hard to produce a luminosity
function with a flat enough faint-end slope, the excess of faint
galaxies perhaps consisting of satellites in massive haloes.

Clearly, matching the luminosity-dependent clustering of galaxies will
continue to be a very stringent test for semi-analytic models. Even if
the models were provided with the correct cosmology as an input,
matching the clustering would still seem to require that the models
predict the correct galaxy population for haloes as a function of
luminosity and mass, rather than predicting quantities which
implicitly average over a range of halo mass, such as the
(unconditional) galaxy luminosity function. Conversely, if the models
were able to correctly capture the trends of luminosity dependent
clustering, it would give us more confidence that they were predicting
realistic galaxy populations on a halo-by-halo basis, and give a
firmer foundation for attempts to constrain cosmology with methods
involving semi-analytic catalogues. Though we bear this in mind, it
seems unrealistic to require a perfect and complete model of galaxy
formation before considering the information it can provide us on
cosmological parameters.

\begin{figure*}
  \begin{center}
    \leavevmode
    \psfig{file=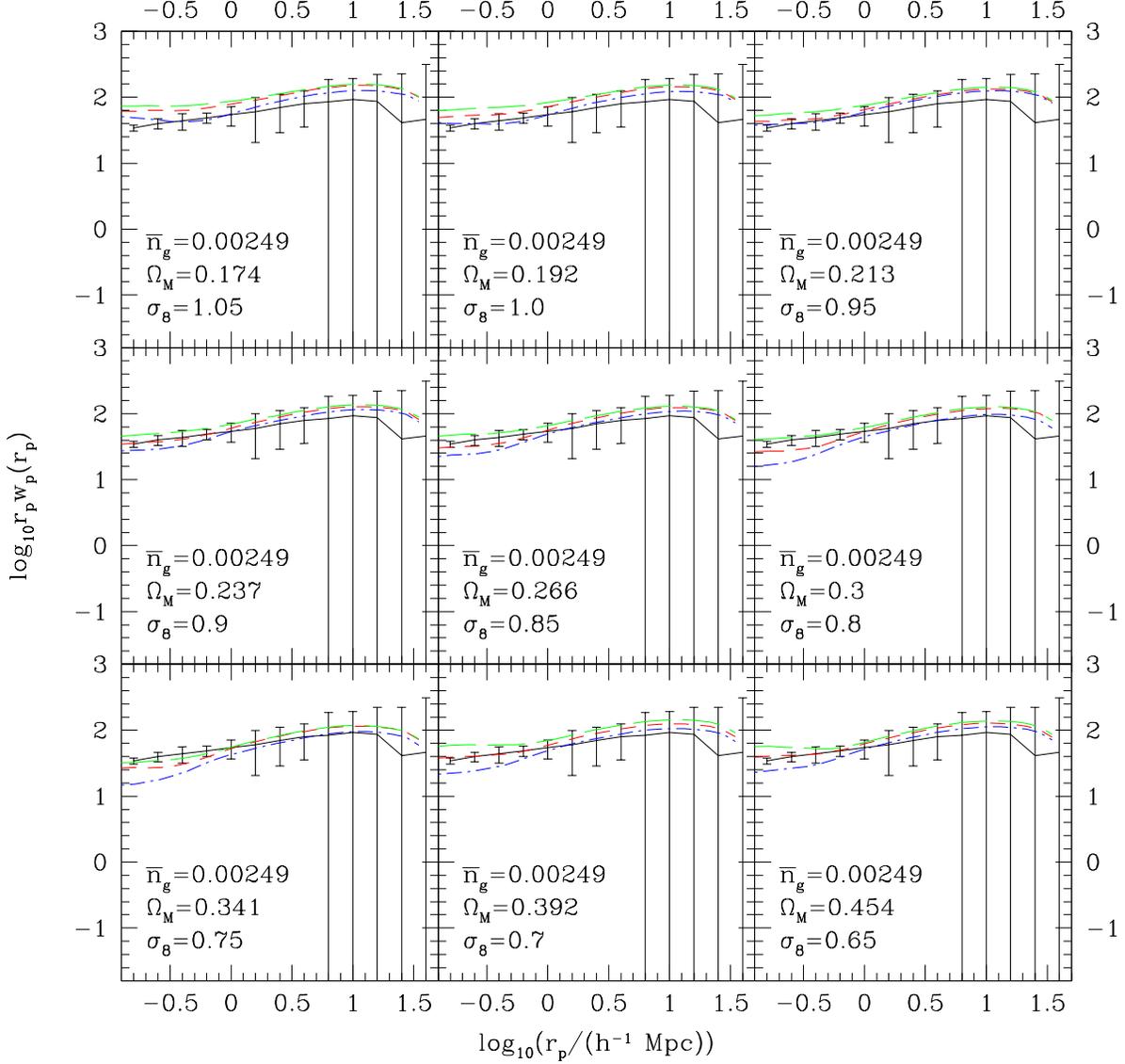,width=16cm}
    \caption{A plot similar to Fig.~\ref{fig:rwgrid1} but for a 2dF
    sample with $M_{b_J}-5\log_{10}h<-18$ and a $b_J$-selected
    \textsc{galform} sample with space density $0.00249\ h^3\
    \mathrm{Mpc}^{-3}$.}\label{fig:rwgrid1_bj}
  \end{center}
\end{figure*}

The models display a minimum in clustering strength at roughly
the space density of the sample we use for our constraint.  We might
therefore expect that using a sample of different space density may
yield a lower estimate for $\sigma_8$ than the sample we have used
above. In fact when we attempt to constrain $\sigma_8$ using different
samples, we obtain high values of $\chi^2$ for all cosmologies, even
for those which appear by eye to be acceptable fits, or which give
reasonable values of $\chi^2$ using only the diagonal elements of the
covariance matrix.  We therefore suspect that for these samples, an
estimate of $\sigma_8$ would be severely affected by noise in the covariance
matrix.  Using only the diagonal elements for samples with
$\bar{n}_\mathrm{g}=0.00031$ or $0.01015\ h^3\ \mathrm{Mpc}^{-3}$
(having $M_\mathrm{r}-5\log_{10}h<-21.5$ and $-19.5$ respectively)
would suggest a slightly lower $\sigma_8$, in the region 0.85--0.9.

One might worry that the supercluster at $z\sim 0.08$ (mentioned
above) affects our constraints. \citet{ZEH05} note, however, that in
their analysis it has no effect on samples fainter than their
$M_\mathrm{r}-5\log_{10}h<-20.0$ sample, while removing it has a very
small effect on brighter samples, producing a negligible drop in
$w_\mathrm{p}(r_\mathrm{p})$. If the drop were larger, then using samples
without this region removed could bias our estimate of $\sigma_8$
upwards. More likely, the supercluster may cause a slight
underestimate of the size of our error bars, since the jackknife
samples used to calculate the covariance matric are smaller than the
supercluster. This prevents the jackknife method from fully capturing
the variance in the density field.

\subsubsection{Constraints from the 2dFGRS}\label{subsubsec:2df}

We have chosen to use the ($r$-band selected) SDSS rather than the
($b_J$-band selected) 2dFGRS for our main constraint on $\sigma_8$,
since the prediction for the luminosity of galaxies in bluer bands
depends more heavily on recent star formation.  It therefore tends to
be more model-dependent than the prediction for redder bands, where
there is a larger dependence on total stellar mass. None the less, the
2dFGRS provides very valuable data on galaxy clustering, and an
accurate galaxy formation model should give constraints on $\sigma_8$
which are consistent between the two datasets. In addition, the
analysis of satellite fractions in the 2dFGRS by \citet{VAN05} suggested, if
somewhat indirectly, that the 2dF data prefer a relatively low
$\sigma_8$. In a similar spirit to the analysis we perform here, this
constraint on $\sigma_8$ came about independently of other
datasets. It may therefore be interesting to see whether our
relatively high value of $\sigma_8$ coming from galaxy clustering data
alone is driven by the data (in which case our estimate for $\sigma_8$
when using 2dFGRS data should be consistent with theirs) or by other
factors. Note, for example, that \citet{PAN05} quote a high preferred
value for $\sigma_8$ in their 2dFGRS clustering analysis -- albeit
concentrating on the three-point function -- though their error bar
extends to low values, $\sigma_8=0.93^{+0.06}_{-0.2}$.

The 2dFGRS clustering data we use are an updated version of the
analysis of \citet{NOR01,NOR02a} (Norberg et al., in prep.). This is
the same dataset as used by \citet{TIN07} in their study of the
luminosity dependence of the galaxy pairwise velocity dispersion. We
compare the sample with $M_{b_J}-5\log_{10}h<-18$ to corresponding
catalogues from our models in Fig.~\ref{fig:rwgrid1_bj}. The grid of
models used is the same as for Fig.~\ref{fig:rwgrid1}, but we select
samples using a $b_J$ magnitude threshold so as to match the space
density of the 2dF sample. The threshold is chosen so that the space
density, $0.00249\ h^3\ \mathrm{Mpc}^{-3}$, is similar to that of our
main SDSS sample. The model clustering appears to depend more weakly
on cosmology than for the $r$-selected samples, but there is still a
clear trend and so we would hope still to be able to use these data to
estimate $\sigma_8$.

We calculate $\chi^2$ between the data and the model using a principal
component analysis, again ignoring the errors on the model correlation
functions as we did for the SDSS.  This analysis is performed on the
dimensionless projected correlation function,
$w_\mathrm{p}(r_\mathrm{p})/r_\mathrm{p}$, denoted
$\Xi(\sigma)/\sigma$ by \citet{NOR02a}. We use only the first six
principal components, which account for over 99 per cent of the
variance.  Statistical errors in the estimate of the principal
components dominate the contribution to $\chi^2$ of the less
significant components. This illustrates the problems which would
arise if we instead used the whole covariance matrix, as highlighted
in Section~\ref{subsubsec:lumdep}.

\begin{figure}
  \begin{center}
    \leavevmode
    \psfig{file=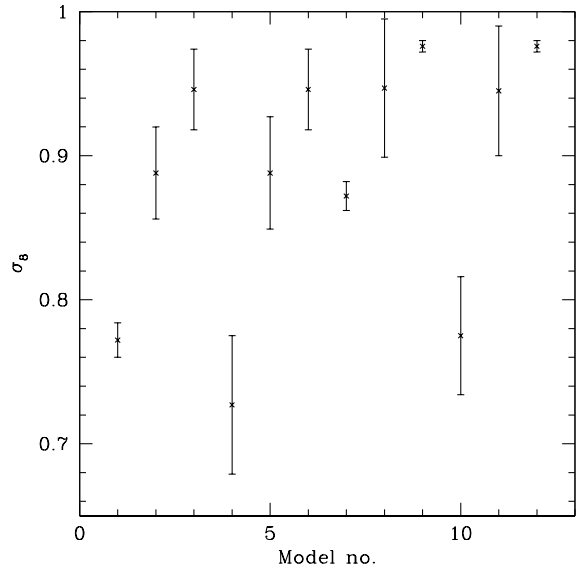,width=8cm}
    \caption{Constraints on $\sigma_8$ for the 2dF sample. The model
    numbering is the same as for Fig.~\ref{fig:bottomline} and is
    given in Table~\ref{tab:modelkey}.}\label{fig:bjpars}
  \end{center}
\end{figure}

The resulting constraints on $\sigma_8$ are given in
Fig.~\ref{fig:bjpars}. The model numbering is the same as for
Fig.~\ref{fig:bottomline}, and is given in
Table~\ref{tab:modelkey}. Noting the change in axis scale from
Fig.~\ref{fig:bottomline}, we see that the statistical error on
$\sigma_8$ from the 2dF sample is comparable to that from the
SDSS. While most of the grids of models yield $\sigma_8$ estimates
similar to those obtained from the SDSS (if perhaps a little lower),
there are several model grids which give significantly lower values of
$\sigma_8$ for the $b_J$-selected samples than they did for the
$r$-selected ones. In fact, these grids (numbers 1, 4 and 10) all use
the C2000hib model. Our results therefore suggest that in this model
the blue galaxies are more clustered than in the others, and hence
lower dark matter clustering is required to match the observational
result. This may be because the feedback excessively reddens isolated
galaxies, leaving a larger proportion of the bluer galaxies in more
massive, more clustered haloes.  In any case it supports the idea that
the clustering of model galaxies selected in bluer wavebands may be
more dependent on the semi-analytic prescription.

\section{Discussion}

As we note in Section~\ref{subsec:const}, the constraints from the 12
different sets of catalogues in Fig.~\ref{fig:bottomline} are not
independent, since the underlying $N$-body simulations in each case
were seeded with the same phases.  This does, though, mean that we can
use the scatter between the catalogues to estimate the systematic
error in the constraint arising from our choice of semi-analytic
prescription.  While we only have three different models (along with
variants in which we do not tweak the parameters to match the $r$-band
luminosity function), we can see that they differ quite strongly in
the luminosity dependence (Fig.~\ref{fig:rwgrid1}) and colour
dependence (Fig.~\ref{fig:bjpars}) of their clustering. They may,
then, be representative of the scatter we can expect between
\textsc{galform} models that fit the $r$-band luminosity function and
the primary constraints listed in Section~\ref{subsec:sams}.  From the
range of $\sim 0.07$ in the value of the best-fitting $\sigma_8$
between sets of catalogues, we estimate a systematic error from this
source of $\pm 0.04$. The average size of the statistical error bars
among the catalogues for which the best-fitting $\sigma_8$ was a good
fit in a $\chi^2$ sense suggests a statistical error of, again $\pm
0.04$.  Adding these in errors in quadrature to the mean of the
best-fitting values in these catalogues gives a final figure of
$\sigma_8=0.97\pm 0.06$ for the $r$-band samples.

Within the scope of the parameter variations we investigated, if
we assume that $\sigma_8=0.8$ then none of our models gives us a good
fit to the SDSS clustering over the full range of scales.  This does
not preclude the possibility that models that do not fit the
luminosity function, or that include different physics to our
particular semi-analytic model, may achieve such a fit.

Our value for $\sigma_8$ is clearly at odds with the most striking
recent measurement, from the three-year WMAP data.  Using those data
alone, \citet{SPE07} quote $\sigma_8=0.761^{+0.049}_{-0.048}$ for
flat, power-law $\Lambda$CDM, and this value is not significantly
increased (though the error bars tighten) when the data are analysed
jointly with galaxy clustering or supernova data. There is, though,
some tension between the WMAP result and results from weak lensing
surveys, which provide rather complementary parameter constraints
\citep{TER05}. \citeauthor{SPE07}'s joint analysis of WMAP and the
CFHTLS lensing survey \citep{HOE06,SEM06} pulls their estimate up to
$\sigma_8=0.827^{+0.026}_{-0.025}$, with the lensing data alone
favouring even higher values. \citet{BEN07} have combined data from
the CFHTLS and other surveys to give
$\sigma_8(\Omega_\mathrm{M}/0.24)^{0.59} = 0.84\pm
0.07$. Lyman-$\alpha$ forest data can be used to constrain the power
spectrum; \citet{SEL05a} quote $\sigma_8=0.90\pm 0.03$ (reducing to
0.84 incorporating the new constraints on reionization from the
three-year WMAP data). Measurements of cluster abundance have
frequently been used to constrain $\sigma_8$, but provide a very wide
range of estimates because of the difficuly in relating the properties
of an observed cluster to its mass \citep[e.g.,][]{RAS05}. Recent
estimates are, though, consistent with the WMAP determination of
$\sigma_8$ \citep[e.g.,][]{PIE03}.

The overall picture of the value of $\sigma_8$ from other methods is
therefore a little confusing, but even the highest recent estimates
are only marginally consistent with ours. A possible source of
tension between our constraints and those from WMAP is that we have
assumed a spectral index $n_\mathrm{s}=1$, while the best-fitting WMAP
$\sigma_8$ is quoted for their best-fitting $n_\mathrm{s}$ of
approximately $0.95$. As one can see from the lower right panel of
figure~10 of \citet{SPE07}, the constraints on these two parameters are
correlated. Even so, increasing $n_\mathrm{s}$ to unity would only
correspond to an increase in $\sigma_8$ of $~0.1$, which is not enough
to eliminate the discrepancy with our result. A further complication
is that the size of $n_\mathrm{s}$ is not the only difference in the
initial $P(k)$ between the WMAP three-year constraints and our
model.  Though ideally one might like to repeat our tests using the
power spectrum shape inferred by \citet{SPE07}, we note that
despite some parameter changes from the first-year WMAP constraints,
the net change in the power spectrum is small. Furthermore, other
datasets that are less dependent on $n_\mathrm{s}$ also prefer a lower
$\sigma_8$ than we do, so even were we to infer a higher $\sigma_8$
from the CMB than currently favoured, our problems would not be solved.

As far as using galaxy data alone goes, methods involving higher-order
correlations, particularly the three-point correlation function (or
its Fourier counterpart the bispectrum) are promising, for example
because of their ability to constrain galaxy bias. The addition of
dynamical information, for example redshift space distortions or the
pairwise velocity dispersion (PVD) can also help constrain
cosmological parameters and galaxy bias. An analysis including PVD
information in the conditional luminosity function (CLF) approach, by
\citet{YAN04,YAN05}, suggested relatively low values of $\sigma_8$,
though this was inferred from their models with high $\sigma_8$ since
low-$\sigma_8$ simulations were not explicitly analysed.  Results
from halo occupation (HOD) modelling, an approach perhaps more akin
to ours -- for example the analysis of the cluster mass-to-light ratio
by \citet{TIN05} -- have tended also to favour a low
$\sigma_8$. \citet{ZHE07}, and references therein, provide a detailed
explanation of HOD modelling and its use in constraining parameters
using galaxy data alone.

An alternative approach using the 2dFGRS is employed by
\citet{VAN05}. They study the abundance and radial distribution of
satellite galaxies within the CLF framework, using mock galaxy
catalogues produced by a semi-analytic code to calibrate their
model. This calibration quantifies the impact of the inevitable
imperfections in the halo finder that lead to satellite galaxies being
spuriously identified as central galaxies of separate haloes, and vice
versa. It also accounts for incompleteness effects in the
2dFGRS. Their results are consistent with other CLF analyses in
suggesting that simultaneously matching the observed cluster
mass-to-light ratio and the fraction of satellite galaxies in the
2dFGRS requires a low value of $\sigma_8$, lowering the abuundance of
very massive haloes with a great number of satellites. Again, they do
not directly construct mock galaxy redshift surveys for a low
$\sigma_8$ model, but using the same calibration parameters as for
their $\sigma_8=0.9$ model leads them to believe that adopting
$\sigma_8=0.7$ provides a better fit to the data.

The parameters of the conditional luminosity functions which
\citet{VAN05} use to fit the 2dFGRS data for low and high $\sigma_8$
are tabulated in their paper. We have used these parameters to
construct the corresponding mean occupation functions -- that is, the
mean number of galaxies in a halo of given mass -- then used these
functions to populate the Millennium Simulation, and outputs of our
Run~1 having $\sigma_8=0.7$ and $\sigma_8=0.9$. This is done as
follows: for each halo in the simulation, we look up the mean number
of galaxies in a halo of this mass, $\langle N(M)\rangle$. If $\langle
N\rangle\leq 1$ the halo receives either a single, central galaxy
(with probability $\langle N\rangle$) or no galaxies at all (with
probability $1-\langle N\rangle$). If $\langle N\rangle >1$ then the
halo receives a central galaxy, plus a number of satellite galaxies
drawn from a Poisson distribution with mean $\langle N\rangle -1$. The
use of these probability distributions follows the work of
\citet{KRA04} and \citet{ZHE05}; in addition, the halo occupation
distribution of \textsc{galform} galaxies in our models is
consistent with this scheme. Once we know the number of galaxies in a
given halo, we then place the galaxies according to the scheme
described in Section~\ref{subsec:galplace}.

We find that the mean occupation functions look reasonable, though in
some cases they are not quite monotonic, and do not generally exhibit
so clean a `step function + power law' form as the \textsc{galform}
mean occupation functions. In our approach to populating simulations
with the CLF HODs, we do not assign a luminosity to each galaxy, and
so we construct a different catalogue for each magnitude threshold we
wish to analyse. The space density of these catalogues as a function
of magnitude gives us a cumulative luminosity function, and we have
checked that this matches the 2dFGRS $b_J$-band luminosity function
for the $\sigma_8=0.9$ catalogue, as it should by construction. For
$\sigma_8=0.7$, the CLF catalogues give too steep a faint-end slope of
the luminosity function, so that strictly speaking the CLF is
incorrect, though this should not be a concern for galaxies of the
luminosity we use for our clustering analysis. We also find that the
CLF produces clustering consistent with our $b_J$-selected
\textsc{galform} samples.

The similar clustering in the \textsc{galform} and CLF catalogues
raises the question of what drives the difference in preferred
$\sigma_8$. The similarity between values from our SDSS and 2dF
analyses suggests it is not driven entirely by the data. Moreover, it
seems at odds with the conclusions of \citet*{TIN06}, who show that in
their HOD model the projected correlation function tightly constrains
the satellite fraction. The answer may lie in the fact that their
parametrized HOD, and our semi-analytic HOD, are unable to match the
form of the mean occupation functions produced by the CLF approach, in
which the parametrizations adopted for different parts of the CLF are
a few steps removed from HOD parameters.

It would be exciting to conclude that our results support a real
difference between low-redshift estimates of $\sigma_8$ (e.g. from
weak lensing) and estimates using CMB data, and that this indicates
something about, say, evolving dark energy. Other analyses find lower
values, though, and there are still one or two concerns about our
constraints. The mean occupation functions from our high
$\Omega_\mathrm{M}$, low $\sigma_8$ \textsc{galform} runs tend to be
more ragged, perhaps indicating a difficulty with our modelling in
these cosmologies.  While we have checked that
galaxy samples selected in a different waveband give similar results,
the anomalous C2000hib model gives some cause for concern.  Similarly,
while samples with a different magnitude threshold appear to give
consistent results, the luminosity dependence of clustering differs
between models. These samples also have smaller effective volume which
seems to give rise to problems in the error analysis. 

As we hoped, a large part of our constraint comes from the
intermediate-scale clustering for which halo-based models are most
necessary, but this is affected by the scheme for placing galaxies
within haloes. The largest difference between our high and low
$\sigma_8$ models (and between our low $\sigma_8$ models and the data)
is manifested at small scales.  The radial distribution of satellite
galaxies within haloes is again different in the \textsc{galform}
model of \citet{BOW06} which uses $N$-body substructure data, and this
is clearly an area of galaxy formation modelling which requires
further attention. We have repeated our analysis of the SDSS
data using only points with $r_\mathrm{p}>2\ h^{-1}\ Mpc$. While the
best-fitting value of $\sigma_8$ decreases by approximately six
per cent to $\sim 0.91$, the error bars approximately double in size,
illustrating the importance of making use of smaller scales.

While bearing the above caveats in mind, we would like to emphasize
the tight constraints available in principle using our technique, and
to note that the results presented here are consistent (if marginally)
with some other analyses which use only low-redshift data. If the
tightness of the constraints seems surprising, consider firstly that in the
absence of uncertainty about galaxy bias, the constraints would be
extremely tight as the amplitude of the correlation function is very
well determined. Secondly, by comparing model galaxies to real
galaxies of the same abundance, we factor out most of the dependence
on the details of the semi-analytic model.  Any change that makes
galaxies monotonically brighter or fainter will have no effect on our
comparison between models and data.  Thirdly, the number of galaxies
assigned to each halo in the semi-analytic model is principally
determined by the merger history of that halo, and this is well
described by the extended Press-Schechter theory.  Hence this is not a
major uncertainty in the semi-analytic models: unlike purely
statistical descriptions as provided by the HOD or CLF, our
semi-analytic models do not have the freedom to define arbitrary
HODs. These are instead largely determined by the merger trees.

We also remark that other constraints using galaxy data alone, which
at first sight seem inconsistent with ours, use different techniques
or different data or both. These inconsistencies arise even though
galaxy formation models can reproduce statistics which implicitly
average over a range of halo mass -- such as the unconditional galaxy
luminosity function -- reasonably well. This illustrates that matching
the luminosity- and colour-dependence of clustering, at small and
large scales, is an important and very stringent requirement on galaxy
formation models.

\section{Conclusions}

We have compared the SDSS projected two-point correlation function at
a galaxy space density $\bar{n}_\mathrm{g}=0.00308\ h^3\
\mathrm{Mpc}^{-3}$ to a suite of populated simulations generated using
the $N$-body code \textsc{gadget2} and the semi-analytic code
\textsc{galform}. Because we require $N$-body data in a great number
of different cosmologies, we have relabelled and rescaled some
simulation outputs using the techniques of \citet{ZHE02} to avoid the need to run a full simulation for each
cosmology in our grid. The galaxy catalogues are self-consistent,
\textsc{galform} being run afresh for each cosmology we study.

We have attempted to estimate the systematic error in our value of
$\sigma_8$ due to the particular choice of semi-analytic model by
running three different \textsc{galform} variants in each
cosmology. For each of these variants we generate a catalogue in which
the \textsc{galform} parameters are adjusted to match the SDSS
$^{0.1}r$-band luminosity function and a catalogue in which the
parameters take the same values as they take in the
$(\Omega_\mathrm{M},\sigma_8)=(0.3,0.8)$ cosmology. We obtain the
result $\sigma_8=0.97\pm 0.04$(statistical) $\pm 0.04$
(systematic). This constraint is impressively tight, given we have
attempted to narrow the range of assumptions we require to produce an
estimate of $\sigma_8$ by using only one well understood, low redshift
dataset. By choosing grids of cosmologies which lie on
cluster-normalized curves,
$\sigma_8\Omega_\mathrm{M}^{0.5}=\mathrm{const.}$ we have shown that
the degeneracies inherent in our approach are different to those
inherent in cosmic shear measurements, which provide an important low
redshift constraint in $\Omega_\mathrm{M}$ and $\sigma_8$. In fact our
method gives an almost pure constraint on $\sigma_8$. We have shown
that in our model, halo assembly bias does not severely affect our
constraint on $\sigma_8$, though this may not be universally the case
for other semi-analytic codes. If it were not the case, we would
expect it to bias our estimate of $\sigma_8$ high, since failing to
account for halo assembly bias tends to lower the amplitude of model
correlation functions, requiring an increased $\sigma_8$ in the model
to compensate.

We obtain similar values for $\sigma_8$ if we use samples with a lower
or higher galaxy space density, but the error analysis is less
secure. We also obtain similar values using a principal component
analysis of 2dFGRS data, for a sample of similar space density to that
used for our primary constraint. We note, though, that the clustering
of galaxies selected in bluer wavebands appears to be more
model-dependent, as one might expect.

Our estimate of $\sigma_8$ looks high compared to the values obtained
by many recent measurements, in particular those from the WMAP
experiment. While we note that this tension has some interesting
consequences if it persists, we have also pointed out how apparent
inconsistencies between our results and other low redshift constraints
may arise.  Small and intermediate scales in the correlation functions
contribute strongly to $\chi^2$ and hence to our constraint on
$\sigma_8$, and yet are not as well understood as the large
scales. This is clearly an area where further modelling effort is
required. Moreover, we will not be completely assured that
semi-analytic models capture the phenomenology of the galaxy
population sufficiently well for high precision cosmological
constraints until they are able to match the observed colour- and
luminosity-dependent clustering of galaxies. The models need to be
able to reproduce the properties of the observed galaxy population on
a halo-by-halo basis, not just the properties averaged spatially or
over luminosity. This implies that if cosmological parameters can be
tightly constrained by other techniques, measurements of galaxy
clustering will continue to provide stringent tests for models of
galaxy formation.

\section*{Acknowledgements}

Most of the work for this paper was undertaken while GH was in receipt
of a PhD studentship from the Particle Physics and Astronomy Research
Council. Thanks to Peder Norberg for providing us with his principal
component analysis of the 2dFGRS data, and his help with the use
thereof. David Weinberg helped to initiate and plan this project, and
provided the SDSS data on which our main analysis is based.

\bibliography{allbib}
\end{document}